\documentclass[10pt,aps,amsmath,amssymb,twocolumn,prl,superscriptaddress,nofootinbib
]{revtex4-2}

\usepackage[utf8]{inputenc}
\usepackage[T1]{fontenc} 
\usepackage{times}  
\usepackage{CJKutf8}

\usepackage{multirow}
\usepackage{booktabs}   
\usepackage{array}   
\usepackage{graphicx}       
\usepackage{tikz}
\usetikzlibrary{shapes.geometric} 
\usepackage{hhline}
\usepackage{mdframed}
\usepackage{microtype}

\usepackage{amsmath}
\usepackage{amssymb}
\usepackage{amsthm}
\usepackage{mathtools}
\usepackage{physics}
\usepackage{braket}

\usepackage{pifont}
\newcommand{\cmark}{\ding{51}}
\newcommand{\xmark}{\ding{55}}

\usepackage{xurl}
\usepackage[
hyperindex,
breaklinks,
colorlinks=true,
citecolor=teal,
urlcolor=blue]{hyperref}
\usepackage[capitalize]{cleveref}

\usepackage{soul}

\theoremstyle{plain}
\newtheorem{theorem}{Theorem}
\newtheorem{lemma}[theorem]{Lemma}

\newtheorem{result}[theorem]{Result}

\theoremstyle{definition}

\newcommand{\hs}{\mathcal{H}}
\newcommand{\C}{\mathbb{C}}

\renewcommand{\r}{\right}
\renewcommand{\l}{\left}
\newcommand{\ten}{\otimes}
\newcommand{\R}{\mathcal{R}}

\usepackage{ulem}
\usepackage{tabularx}
\newcolumntype{C}{>{\centering\arraybackslash}X}
\newcolumntype{M}[1]{>{\centering\arraybackslash}m{#1}}

\begin{document}

\title{Large Parts are Generically Entangled Across All Cuts}

\author{Mu-En Liu}
\email{l26121028@gs.ncku.edu.tw}
\affiliation{Department of Physics and Center for Quantum Frontiers of Research \& Technology (QFort), National Cheng Kung University, Tainan 701, Taiwan}

\author{Kai-Siang Chen}
\affiliation{Department of Physics and Center for Quantum Frontiers of Research \& Technology (QFort), National Cheng Kung University, Tainan 701, Taiwan}

\author{Chung-Yun Hsieh}
\affiliation{H. H. Wills Physics Laboratory, University of Bristol, Tyndall Avenue, Bristol, BS8 1TL, UK}

\author{Gelo Noel M. Tabia}
\affiliation{Department of Physics and Center for Quantum Frontiers of Research \& Technology (QFort), National Cheng Kung University, Tainan 701, Taiwan}
\affiliation{Foxconn Quantum Computing Research Center, Taipei 114, Taiwan}
\affiliation{Physics Division, National Center for Theoretical Sciences, Taipei 106319, Taiwan}

\author{Yeong-Cherng Liang}
\email{ycliang@mail.ncku.edu.tw}
\affiliation{Department of Physics and Center for Quantum Frontiers of Research \& Technology (QFort), National Cheng Kung University, Tainan 701, Taiwan}
\affiliation{Physics Division, National Center for Theoretical Sciences, Taipei 106319, Taiwan}
\affiliation{Perimeter Institute for Theoretical Physics, Waterloo, Ontario, Canada, N2L 2Y5}
\email{ycliang@mail.ncku.edu.tw}

\begin{abstract}
    Generic high-dimensional bipartite pure states are overwhelmingly likely to be highly entangled.
    Remarkably, this ubiquitous phenomenon can already arise in finite-dimensional systems.
    However, unlike the bipartite setting, the entanglement of {\em generic multipartite} pure states, and specifically their multipartite {\em marginals}, is far less understood.
    Here, we show that sufficiently large marginals of generic multipartite pure states, accounting for approximately half or more of the subsystems, are {\em entangled} across {\em all} bipartitions.
    These pure states are thus robust to losses in entanglement distribution and potentially useful for quantum information protocols where the flexibility in the collaboration among subsets of clients is desirable.
    We further show that these entangled marginals are not only shareable in closed systems, but must also induce entanglement in other marginals when some mild dimension constraints are satisfied, i.e.,  entanglement transitivity is a generic feature of various many-body closed systems. 
    We further observe numerically that the genericity of (1) entangled marginals, (2) unique global compatibility, and (3) entanglement transitivity may also hold beyond the analytically established dimension constraints,  which may be of independent interest.
\end{abstract}

\date{\today}
\maketitle

{\bf\em Introduction---}Characterizing various properties of quantum states, such as entanglement~\cite{Horodecki09_EntReview, Guhne08_EntReview}, remains a challenging task in higher-dimensional systems, where many complex phenomena take place that defy simple laws.
Surprisingly, as the dimension of the space grows, most quantum states turn out to share some common features.
For example, most pure bipartite states are known to be highly entangled~\cite{Hayden06_GenericEnt}.
Thus, although it is hard to characterize all quantum states, it is still possible to tell if {\em most} states possess certain properties.
The knowledge of how generic states behave also has physical implications in thermodynamics~\cite{Popescu06_Ent_StatMech, Muller15_Typicality, Goold16_review_QIandThermo}, quantum computing~\cite{Gross09_UsefulEnt}, black hole physics~\cite{Hayden07_BlackHoleMirror}, and many more~\cite{Ma24_PRU}.
Random matrix theory, which serves as a powerful tool to study generic states and many other phenomena in quantum information, has yielded many fruitful results~\cite{Collins15_random_matrix, Aubrun17_AliceandBob, Mele24_HaarReview},
including disproving the additivity conjecture~\cite{Hastings09_Additivity, Aubrun10_Additivity}, and making partial progress~\cite{Collins18_PPT} on the positive-partial-transpose (PPT)~\cite{Peres96_PPT} square conjecture~\cite{baeuml2010thesis, Ruskai12}.

\begin{figure}[h!]
  \centering
  \includegraphics[width=0.4\textwidth]{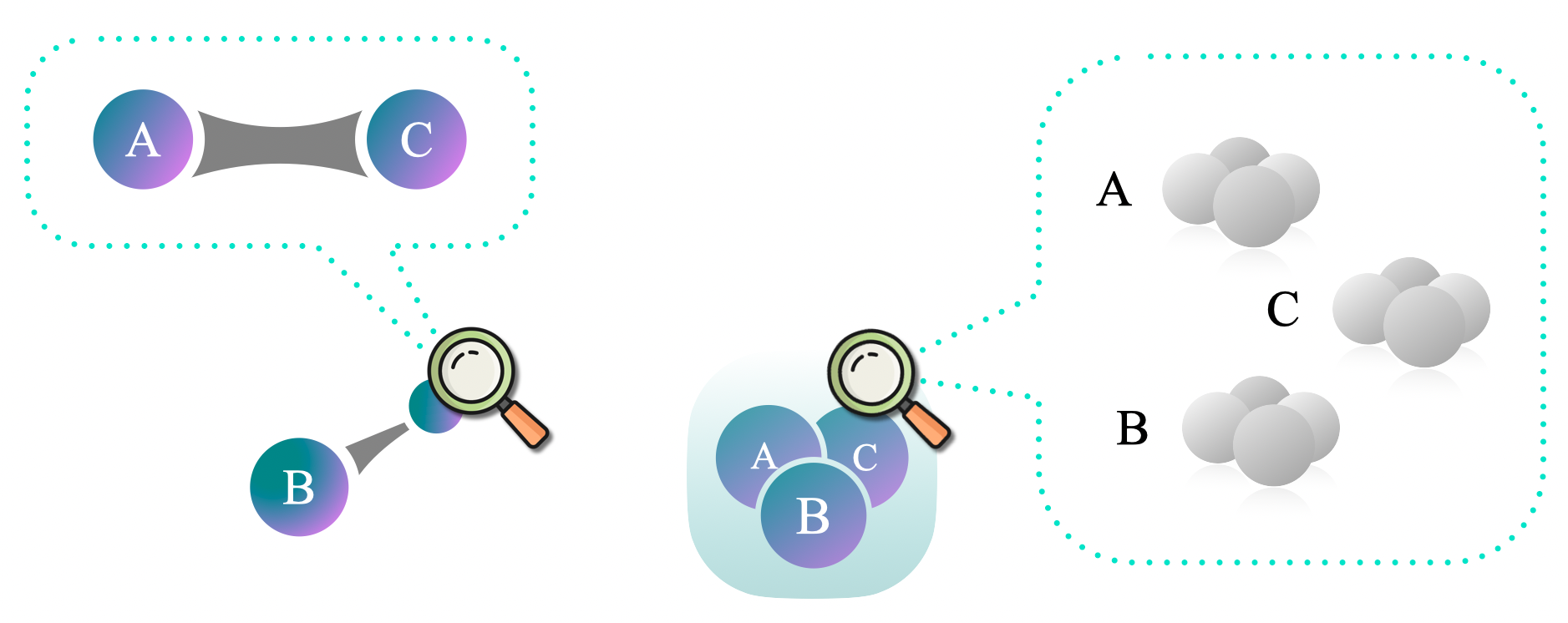}
  \caption{\small
    (Left) When a bipartite pure state includes a subsystem that is itself also formed by two subsystems (labeled with A and C here), the overall pure state can be reinterpreted as a tripartite state on systems A, B, and C.
    (Right) This tripartite state may arise from many even smaller constituents, making the AC subsystem a many-body marginal.
    We investigate whether such a many-body subsystem is entangled across arbitrary bipartitions of its components.
  }
  \label{fig:ent_marg}
\end{figure}

Most bipartite pure states in the limit of large (unbalanced) local dimensions are known to be highly entangled~\cite{Hayden06_GenericEnt}.
A natural question is whether their subsystems, when formed by even smaller bipartite systems (see~\cref{fig:ent_marg}), also possess entanglement.
Indeed, since we often have access only to parts of the total system (whose physical states are called the marginal states of the global system, or simply {\em marginals}), it is crucial to understand whether the marginals of most pure states are entangled.
To this end, it is known that when the bipartite marginals arising from high-dimensional pure states are relatively {\em small} (in their combined state space dimension), they are overwhelmingly likely~\footnote{
By overwhelmingly likely, or very likely, we mean that the probability is bounded away from unity by a term that decays exponentially, with some parameters (e.g., dimensions).}
separable~\cite{Hayden06_GenericEnt, Aubrun13_EntThreshold}. 
In contrast, relatively {\em large} parts are almost surely entangled~\citep[Proposition 7.1 (ii)]{Aubrun13_EntThreshold} (see also~\cite{Walgate08_RandomSubspace, Ruskai09_GenericEnt})~\footnote{ 
Aubrun {\em et al.}~\cite{Aubrun13_EntThreshold, Aubrun12_PRA} showed that typical marginal entanglement overwhelmingly likely follows an {\em all-or-none} behavior: given marginals of the same size, the vast majority will either be {\em all} entangled or {\em all} separable.
Such all-or-none behavior also holds for several other supersets of the separable set~\cite{Aubrun12_PRA, Aubrun12_PPT, Aubrun12_realignment_generic_state, Jivulescu14_reduction_generic_state, Lancien16_extendibility_generic_state}, as summarized in~\cite{Collins15_random_matrix}.}, i.e., the probability of finding a separable one by chance is {\em exactly} zero.
Specifically, a bipartite marginal $\rho_{AC}$ induced from a tripartite Haar-random pure state $\ket{\psi}_{ABC}$ is almost surely entangled if the local finite dimensions satisfy an elegantly simple relation~\cite{Walgate08_RandomSubspace, Aubrun13_EntThreshold, Ruskai09_GenericEnt, Lockhart02_EntLowRank, Harrow13_SymmetricSubspace}.

What if the individual subsystems A, B, C, alluded to above, are themselves decomposable, e.g., formed by many constituents (see~\cref{fig:ent_marg})?
Based on the discussion above, one may rightfully expect that relatively small many-body marginals are overwhelmingly likely to be {\em fully} separable~\cite{Aubrun13_EntThreshold}, and sufficiently large many-body marginals of generic pure states are also very likely to be entangled across all cuts~\citep[Corollary VI.2]{Hayden06_GenericEnt},
or even {\em almost surely} entangled with respect to certain bipartitions~\citep[Proposition 7.1(ii)]{Aubrun13_EntThreshold}
(see also~\cite{Kendon02_MarginalEnt}).
In this work, we sharpen these intuitions by showing that the many-body marginals of a {\em generic} random pure state in finite-dimensional systems are entangled with respect to {\em all bipartitions},
provided that they are slightly larger than half of the pure state to which they belong.

{\bf\em Preliminary notions---}Consider a tripartite Haar-random pure state $\ket{\psi}_{ABC} \in \C^{d_A} \ten \C^{d_B} \ten \C^{d_C}$, and its bipartite marginal $\rho_{AC} \coloneqq \tr_B(\ketbra{\psi}_{ABC})$.
The subscript $X=A,B,C$ beside the state $\ket{\psi}_X, \rho_X$ denotes the (sub)system with which the state is associated.
$d_X$ denotes the dimension of the subsystem $X$, and $\tr_B$ denotes the partial trace over the subsystem B.
Here, the local dimensions are assumed to be finite, i.e., $d_A, d_B, d_C < \infty$.
A Haar-random pure state~\citep[Definition 4.1]{Collins15_random_matrix} is one that is uniformly sampled from all possible pure states and is also called a {\em generic} state.
By generalizing~\citep[Proposition~7.1 (ii)]{Aubrun13_EntThreshold}, one can obtain the following result (see also~\citep[Theorem~1]{Lockhart02_EntLowRank}).
\begin{lemma}
\label{lem: entmarg}
    $\rho_{AC}$ is almost surely entangled if the dimensions of the generic tripartite pure state satisfy
        $
	     d_B \leq (d_A-1)(d_C-1).
	$
\end{lemma}
\noindent 
Here, by saying ``almost surely entangled'', or ``generically entangled", it formally means that for uniformly sampled $\ket{\psi}_{ABC}$, those with marginals in AC that are separable form a measure-zero set.
Namely, \cref{lem: entmarg} implies that the marginal, $\rho_{AC}$, of a finite-dimensional randomly sampled pure state, $\ket{\psi}_{ABC}$, are guaranteed to be entangled, provided that they are large enough, $d_Ad_C \geq d_A+d_B+d_C-1$.
This implies that, in a tripartite closed system with equal finite local dimension $d_A = d_B = d_C \geq 3$~\footnote{The case of three qubits was covered separately in~\citep[Proposition~7.1 (iii)]{Aubrun13_EntThreshold}.}, all two-body subsystems are generically entangled.
Hence, it is extremely hard to find a two-body separable state in a generic three-body closed system---in fact, it happens with zero probability.
Note that \cref{lem: entmarg} identifies strictly more entangled marginals of generic tripartite pure states than the sufficient condition of distillability---also expressible as an inequality of $d_A,d_B$, and $d_C$---provided in~\citep[Theorem 1]{Horodecki03_rank2_boundstate}, see~\hyperlink{app:entmarg-remark}{Appendix A} for further details.

For completeness, we provide in~\hyperref[Supplementary Material]{{Supplementary Material I}} the proof of Lemma~\ref{lem: entmarg}.
To shed light on why the Lemma holds, we outline below some key observations.
First, when~\cref{lem: entmarg} holds, the marginal $\rho_{AC}$ of a Haar-random pure state is supported on a subspace void of product states, i.e., the subspace is {\em completely entangled}~\cite{Wallach00_CES, Parthasarathy04_CES}. 
Then, according to the range criterion~\cite{Horodecki97_RangeCriterion}, which states that a separable state must be supported on a subspace containing some product states, one immediately concludes that $\rho_{AC}$ is entangled.

{\bf\em Large enough marginals are entangled across all cuts in random pure states---}Building on Lemma~\ref{lem: entmarg}, we prove our main result, which shows that the multipartite marginals of a generic pure state are almost surely entangled with respect to {\em all} bipartitions once they are slightly larger than half of the entire system 
(see \hyperlink{app:multi-marg}{Appendix B} for a proof).
\begin{theorem}
\label{thm: multipartite entmarg}
    The $k$-qudit marginals of an $N$-qudit generic pure state (with $N \geq 3$ and $N \geq k$) are almost surely entangled with respect to the bipartition of $(\C^d)^{\otimes m} \otimes (\C^d)^{\otimes k-m}$ for every $ 1\leq m \leq k-1$ (i.e., they are entangled with respect to any bipartition), if 
    \begin{equation}
    \label{eq: multipartite entmarg dimension}
        \begin{cases}
            \ k \geq \frac{N}{2} + 1 &\text{when } d=2 \text{ and } m = 1, k-1\\
            \ k > \frac{N}{2} &\text{otherwise}.
        \end{cases}
    \end{equation}
\end{theorem}

\begin{figure}[h]
  \centering
  \includegraphics[width=0.45\textwidth]{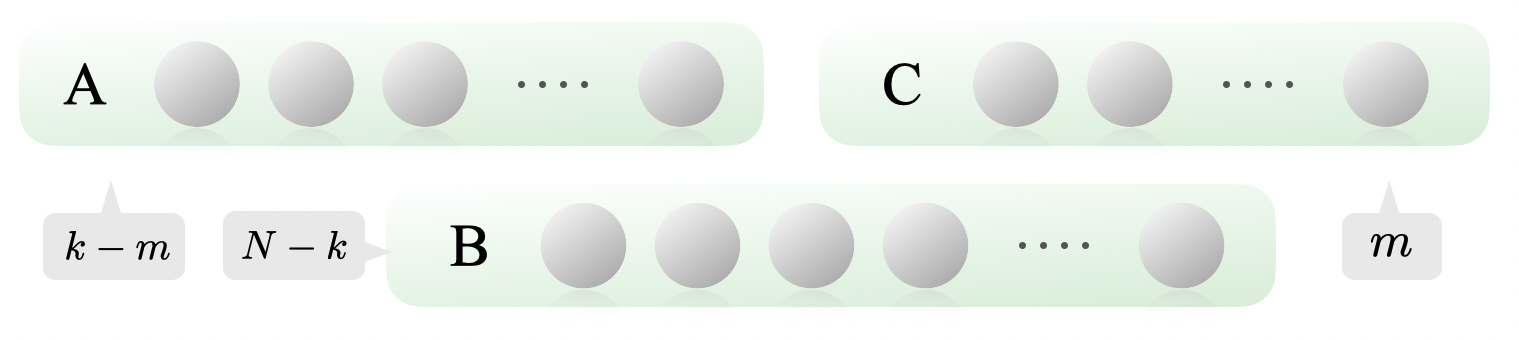}
  \caption{\small
    Illustration of an $N$-qudit generic pure state divided into three blocks labeled with A, B, and C, which correspond to subsystems of $k-m, N-k,$ and $m$ qudits, respectively.
    Each circle represents a qudit. \cref{thm: multipartite entmarg} shows that when $d\ge 2$ and \cref{eq: multipartite entmarg dimension} holds, the $k$-qudit marginal (collectively labeled by AC) is entangled for all cuts (i.e., the separation of the $k$ qudits into two groups can be arbitrary).
    }
  \label{fig:multi_ent_marg}
\end{figure}

Consequently, large enough marginals of an $N$-qudit Haar-random pure state are entangled.
In other words, Haar-random pure states retain entanglement despite the loss of nearly {\em half of} their subsystems, cf.~\cite{Neven18_particle_loss} and references therein.

{\bf\em Haar-random pure states as a resource in loss-tolerant entanglement distribution---}Interestingly, \cref{thm: multipartite entmarg} implies that Haar-random pure states, which are already useful in various quantum information protocols~\cite{Harrow04_superdensecoding, Hayden04_randomize_states, Chen25_SingleHaar}, can serve as a {\em resource} in distributing entanglement~\cite{Perseguers13_EntDistribution} with a high loss tolerance.
To illustrate this, consider the following quantum internet setting with $k$ agents, consisting of $k-1$ clients and a provider.
The provider aims to distribute a $k$-qudit state (with $d\ge3)$, denoted by $\eta^{(k)}$, among all agents (including the provider and all clients), such that $\eta^{(k)}$ is entangled across {\em all} bipartitions---namely, among all possible cuts dividing the agents into two groups, there is always entanglement between these two groups. This way of distributing entanglement is important for, e.g., demonstrating a genuine ``quantum'' internet across all cuts of agents.

Crucially, physically distributing qudits from providers to clients often suffers from {\em particle loss}---for instance, if we encode quantum information in photons, 
having photon loss is unavoidable when the transmission distance is long enough. It is thus vital to determine whether sending {\em more} particles can help achieve the above-mentioned goal.
To illustrate how Haar-random pure states can be a resource, consider a particular scenario where the particle transmission from the provider to each client is subject to 50$\%$ chance of particle loss. 
Loosely speaking, when the provider sends two qudits to a client, that client can, on average, receive one.

To address the issue of particle loss, the provider can start with $(2k-1)$-qudit Haar-random pure states. 
Then, the provider keeps {\em one} qudit (which is assumed to have no particle loss), and sends {\em two qudits} to each client.
In total, we have $2(k-1)+1=2k-1$ qudits.
Now, using \cref{thm: multipartite entmarg} with $d\ge3$ and $N=2k-1$, we obtain $N<2k$, meaning that the $k$ agents (provider plus clients) almost surely shares entanglement across {\em all} bipartitions, provided that {\em each client receives at least one qudit}.
Note that the protocol can be readily adapted so that entanglement can be shared between the provider and {\em each} individual client by allocating more qudits to the provider's side.
Consequently, in this particular scenario,
\begin{center}
{\em
Haar-random pure states are useful for entanglement distribution even when each channel has up to 50$\%$ loss rate.
}
\end{center}
Evidently, the above discussion is only meant to be a toy model aiming to exemplify \cref{thm: multipartite entmarg}'s applicability to loss-tolerant entanglement distribution. 
A more thorough understanding and detailed analysis are desired, which, however, is beyond the scope of this work, and we will leave it for future research.

{\bf\em Haar-random pure states and entanglement sharing game---}Apart from the above implication, \cref{thm: multipartite entmarg} also suggests that Haar-random pure states may be useful for protocols that share the following characteristics.
Here, rather than discussing a specific task, we focus on a high-level description to capture the role of \cref{thm: multipartite entmarg}.
Suppose a task is designated to $N$ clients, which should be completed with some entangled state distributed to them.
The task requires that
\begin{enumerate}
    \item
    \label{characteristics1} 
    At least $\lceil\frac{N}{2}\rceil+1$ clients must collaborate (e.g., in secret sharing~\cite{Hillery99_secret_sharing}) and utilize the shared entanglement among them for the given task.
    \item
    \label{characteristics2} 
    Clients may not know in advance with whom they should collaborate before the state is distributed (i.e., clients cannot form alliances before the protocol begins, cf.~\cite{Liang14_AnonymousNonlocality}).
\end{enumerate}
Here and below,$\lceil \cdot \rceil$ ($\lfloor \cdot \rfloor$) denotes the ceiling (floor) function.
Haar-random pure states thus serve as a natural candidate for tasks with these characteristics.
When an $N$-qudit Haar-random pure state $\ket{\psi} \in (\C^d)^{\ten N}$ is distributed such that each client receives a single qudit,
\cref{thm: multipartite entmarg} guarantees that its $k$-body marginals are almost surely entangled as long as $k \geq \frac{N}{2}+1$. 
This ensures that $\lceil\frac{N}{2}\rceil+1$ or more clients will share entanglement, thereby meeting condition~\ref{characteristics1}. 
Furthermore, condition~\ref{characteristics2} is also satisfied since the state imposes no restrictions on which subset of clients must collaborate;
entanglement is guaranteed for {\em any} subset involving $\lceil\frac{N}{2}\rceil+1$ or more clients, even with respect to any bipartition.

{\bf\em Entanglement transitivity is generic in closed systems---}Another application of Lemma~\ref{lem: entmarg} and~\cref{thm: multipartite entmarg} is that {\em entanglement transitivity}~\cite{Tabia22} is provably generic in closed systems.
We begin by reviewing the definition of entanglement transitivity and sketching the proof of this result.

Consider a tripartite system ABC. 
Two bipartite states $\rho_{AB}$ and $\rho_{BC}$ are {\em compatible} if there exists at least one tripartite state $\eta_{ABC}$ satisfying $\rho_{AB} = \tr_C (\eta_{ABC})$ and $\rho_{BC} = \tr_A (\eta_{ABC})$, i.e., these two bipartite states can be simultaneously realized (via a single global state).
Now, suppose that $\rho_{AB}, \rho_{BC}$ are both entangled and compatible. 
They are said to exhibit {\em entanglement transitivity} in the system AC if for every $\eta_{ABC}$ returning these marginals, $\tr_B(\eta_{ABC})$ {\em must} be entangled in AC.
To illustrate this notion, consider the $W$ state~\cite{Dur00_SLOCC} ${\ket{W}_{ABC} = \frac{1}{\sqrt{3}}(\ket{001}+\ket{010}+\ket{100})_{ABC}}$, whose marginals in AB and BC are entangled,
and furthermore, the only~\cite{Parashar09_wstate} state in AC compatible with both is $\tr_B(\ketbra{W}_{ABC})$.
Thus, the two bipartite marginals of a $W$ state exhibit entanglement transitivity.
More examples can be found in~\cite{Tabia22, Chen24_NonlocalityTrans, Shi25_symmetric_ent_trans}.

\begin{table*}[ht]
  \centering
  \begin{tabularx}{\textwidth}{|M{4cm}|C|M{4cm}|C|}
   \hline
       {\footnotesize \textsc{Reference}}
     & {\footnotesize \textsc{Marginal size threshold}}
     & {\footnotesize \textsc{Probability}}
     & {\footnotesize \textsc{Local dimension constraints}} \\
   \hline
   Hayden \emph{et al.}~\cite{Hayden06_GenericEnt}  
     & $k > N/2 + \mathcal{O}\l(\frac{\log(\log d)}{\log d}\r)$~\footnote{
     The threshold is derived using the entanglement of formation~\cite{Horodecki09_EntReview} as the underlying entanglement measure. 
     } 
     & Very likely
     & $d \ge 3$ \\
   Aubrun \emph{et al.}~\cite{Aubrun13_EntThreshold} 
     & $k > (N-1)/2 - \mathcal{O}\l(\frac{1}{\log d}\r)$~\footnote{
       This follows from the result in~\citep[Section 7.2]{Aubrun13_EntThreshold}.
       For a proof, see~\hyperlink{app:marginal-threshold}{Appendix E}.
       } 
     & Very likely
     & $d \ge 2$ \\
   This work (\cref{thm: multipartite entmarg})
     & $k > N/2\; \l(k \ge N/2+1 \r)$ 
     & Almost surely
     & $d \ge 3$\;($d=2$) \\
   \hline
 \end{tabularx}
  \caption[\small]{
  Summary of results on entanglement across all cuts in marginals induced from generic pure states.
  The table lists the marginal size thresholds above which the $k$-qudit marginals of an $N$-qudit generic pure state are very likely (or almost surely) entangled across all bipartitions.
  $\mathcal{O}(\cdot)$ denotes the standard Big O notation.
  }
  \label{tab:ent_all_cuts}
\end{table*}

Next, we show that the AB and BC marginals of a Haar-random tripartite pure state $\ket{\psi}_{ABC}$---denoted, respectively, by $\rho_{AB}$ and $\rho_{BC}$---generically exhibit entanglement transitivity in AC,
provided that their local dimensions, $d_A, d_B, d_C$, satisfy certain relations.
In other words, entanglement transitivity is generic when $\rho_{ABC}$ describes a closed system, i.e., when it is a pure state and when certain dimension constraints hold.
This may seem surprising, as the problem of characterizing all global states compatible with given marginals---a variant of the quantum (state) marginal problem (see references in~\cite{Fraser22_QMP})---is known to be computationally hard in general~\cite{Liu06_QMP_QMA, Broadbent22_QMP_QMA}.
However, as was first noted by~\cite{Linden02_qubit_uni, Linden02_tripartite_uniqueness}, under specific dimensional constraints, a generic pure state $\ket{\psi}_{ABC}$ is {\em uniquely determined} by its marginals $\rho_{AB}$ and $\rho_{BC}$---i.e., there exists no other state in ABC, pure or mixed, compatible with these two marginals.
In this context, we may summarize the state of the art of such findings as follows (see~\cite{Chen13_tripartite_uniqueness, Huang18_tripartite_uniqueness} and references therein).
\begin{lemma}[Uniqueness~\cite{Chen13_tripartite_uniqueness, Huang18_tripartite_uniqueness}]\label{lem: tripartite uniqueness}
    If $d_B \geq \min\{d_A,d_C\}$, or if $d_B \geq 2$ and $d_A = d_C$, then a generic tripartite pure state $\ket{\psi}_{ABC}$ is uniquely determined by its bipartite marginals $\rho_{AB}, \rho_{BC}$.
\end{lemma}
Consequently, to demonstrate that the entangled marginals $\rho_{AB}, \rho_{BC}$ exhibit entanglement transitivity when any of these dimension constraints hold, 
it suffices to show that $\rho_{AC} = \tr_C(\ketbra{\psi}_{ABC})$ is entangled,
since it is the {\em only} state in AC compatible with both marginals.
Therefore, the formidable task reduces to identifying when the marginal $\rho_{AC}$ of a generic pure state is entangled.
\cref{lem: entmarg} characterizes the conditions under which $\rho_{AB}, \rho_{BC}, \rho_{AC}$ are entangled, 
while~\cref{lem: tripartite uniqueness} ensures uniqueness.
Together, they yield the sufficient conditions for $\rho_{AB}$ and $\rho_{BC}$ to exhibit entanglement transitivity 
(below, we define $d_{\rm min} \coloneqq{\rm min}\{d_A,d_C\}$ and $d_{\rm max} \coloneqq{\rm max}\{d_A,d_C\}$):
\begin{theorem}
\label{thm: ent_trans}
    Let $\rho_{AB}$ and $\rho_{BC}$ be the bipartite marginals of a Haar-random pure state in ABC.
    $\rho_{AB}$ and $\rho_{BC}$ almost surely exhibit entanglement transitivity if one of the following conditions is satisfied:
    \begin{enumerate}
        \item\label{4-1} $d_A = d_B = d_C = 2$,
        \item\label{4-2} $d_A = d_C = d \geq 3$ and
            $1+\frac{d}{d-1} \leq d_B \leq (d-1)^2,$
        \item\label{4-3} $d_{\rm min} 
        \geq 3$ and 
        \begin{equation*}
            {\rm max}\left\{d_{\rm min}, 1 + \dfrac{d_{\rm max}}{d_{\rm min}-1}\right\} \leq d_B \leq (d_A-1)(d_C-1).
        \end{equation*}
    \end{enumerate}
\end{theorem}
See~\hyperlink{app:ent-trans}{Appendix C} for proof.
Note that $d \ge 1+\frac{d}{d-1}$ for all $d\ge 3$. 
Hence, even given the last condition above, the second condition is not redundant.
Importantly, \cref{thm: ent_trans} implies:
\begin{result}
    \label{lab: ent_trans_equal_dim}
    Entanglement transitivity is generic in any tripartite closed system with equal finite local dimension.
\end{result}
This answers a conjecture raised in~\cite{Tabia22} in the affirmative.
Furthermore, it implies that {\em meta-transitivity}~\cite{Tabia22}---the phenomenon where $\rho_{AB}$ and $\rho_{BC}$ are not entangled, yet all compatible $\sigma_{AC}$ are---is a {\em measure-zero phenomenon} when three local dimensions are equal and the global state is pure.
In fact, our numerical results summarized in~\cref{tab:ent_trans_not_specified} in~\hyperlink{app:ent-trans-unspecified}{Appendix F} suggest that generic entanglement transitivity, and more generally, generic entangled marginals and the uniqueness property, also hold {\em beyond} the dimension constraints specified, respectively, in~\cref{thm: ent_trans}, Lemma~\ref{lem: entmarg} and~\cref{lem: tripartite uniqueness}.
In all tested dimensions, marginals of generic pure states seem to always uniquely determine the global state, a phenomenon that may be of independent interest.

Further interesting implications can be drawn from~\cref{thm: ent_trans}.
Consider, e.g., the case when $d_B=3$ and $d_A=d_C=1000$.
Subsystem B is indisputably microscopic, whereas subsystems A and C are relatively macroscopic.
The second condition in~\cref{thm: ent_trans} guarantees that generic pure states with these local dimensions have marginals in AB and BC that exhibit entanglement transitivity in AC. 
Consequently, for such a generic closed system ABC,
the entanglement between microscopic (B) and macroscopic (A, C) parts {\em forces} the two macroscopic subsystems to be almost surely entangled.
Hence, interestingly, the co-existence of two microscopic-macroscopic entangled pairs can imply {\em macroscopic-macroscopic} entanglement in a generic closed system.

So far, we have shown that entanglement is generically transitive in three-body closed systems. 
A natural question is: {\em what about $N$-body closed systems}? 
Clearly, various notions of multipartite entanglement transitivity can be formulated. 
In the following, taking advantage of~\cref{thm: multipartite entmarg}, we focus on the case where both the given and the target marginals are entangled across all bipartitions. Then, similar to the previous approach, we show that
\begin{theorem}[Informal version]
\label{thm: multi_ent_trans}
For a generic $N$-qudit pure state $\ket{\psi} \in (\C^d)^{\otimes N}$:
\begin{itemize}

    \item two specifically chosen $(\lfloor \frac{N}{2} \rfloor + 1)$-body marginals (assuming either $d \geq 3$ or that $N$ is even), or
    
    \item $\lfloor \frac{N}{2} \rfloor$ specifically chosen $(\lceil \frac{N}{2} \rceil + 1)$-body marginals,
    
\end{itemize}
almost surely exhibit entanglement transitivity in all remaining $k$-body marginals with $k$ satisfying~\cref{eq: multipartite entmarg dimension}.
\end{theorem}
See~\hyperlink{app:multi-ent-trans}{Appendix D} for a formal version of the statement and its proof.
To illustrate our result, let $\ket{\psi}_{ABCD}$ be a generic four-qudit pure state.
Two of its $3$-body marginals, \( \{\rho_{ABC}, \rho_{BCD}\}\), cannot be simultaneously compatible with any other $3$-body states other than $\ket{\psi}_{ABCD}$'s remaining $3$-body marginals \( \{\rho_{ACD}, \rho_{ABD}\} \), which are almost surely entangled across all cuts due to both conditions in~\cref{thm: multipartite entmarg}. 
Thus, the marginals \( \{\rho_{ABC}, \rho_{BCD}\}\) exhibit entanglement transitivity in all the remaining three-qudit marginals, as every compatible marginal with the same number of constituents is guaranteed to be entangled across all cuts.

{\bf\em Discussion---}The investigation of properties of generic states has led to the realization that entanglement is ubiquitous in high-dimensional closed systems.  
In this case, most pure states are nearly maximally entangled, and their marginals, formed by even smaller constituents, are also entangled as long as they are large enough---a fact encapsulated in Lemma~\ref{lem: entmarg}, along with other results~\cite{Aubrun13_EntThreshold}. 
More precisely, large separable marginals form a zero-measure set, which is surprising given that separable states are not of measure zero in the entire state space~\cite{Gurvits02_separable_ball}.

In this work, we show in~\cref{thm: multipartite entmarg} that the marginals can even be entangled with respect to {\em every} bipartition as long as their size is slightly larger than half of the global pure state they belong to, cf.~\cref{eq: multipartite entmarg dimension}. 
Naturally, one may wonder if this bound is tight. 
To this end, we note that this bound is consistent with
previous results (organized in~\cref{tab:ent_all_cuts}) in the asymptotic limit of large $d$ (i.e., $d\to\infty$).
Interestingly, our result indicates a distinct threshold for qubit systems (cf.~\cite{Hayden06_GenericEnt}).
Whether the bound is truly different for qubit systems is left for future investigation.

In comparison, entanglement across balanced bipartitions (i.e., $m=k/2$) becomes overwhelmingly likely when $k>\frac{2N}{5}$~\citep[Corollary 2.4]{Aubrun13_EntThreshold}, showing that much smaller marginals suffice if we only demand entanglement across bipartitions of certain sizes. 
Intriguingly, numerical evidence in~\cite{Kendon02_MarginalEnt} suggests that when the marginal size $k$ exceeds $\frac{N}{2}$, the marginal becomes nonpositive under partial transposition (NPPT) across at least one bipartition with high probability.
This indicates that our bound could well be a threshold for NPPT entanglement across at least one cut, which remains to be proved.

Entanglement is known to be monogamous~\cite{Coffman00_EntMonogamy}; it is neither arbitrarily nor infinitely shareable~\cite{Fannes88, Raggio89, Doherty04_symmetric_extension, Doherty14_review_extension}.
The phenomenon of entanglement transitivity~\cite{Tabia22} complements this view: it reveals that certain entangled marginals, in fact, necessitate entanglement in other marginals. 
In Theorem~\ref{thm: ent_trans} (\cref{thm: multi_ent_trans}), we demonstrate that the marginals of generic pure states indeed exhibit (multipartite) entanglement transitivity. Notably, these generic marginals also defy monogamy-type relations~\cite{Lancien16_ent_measures}.

For our results on multipartite entanglement transitivity, we have focused on the notion of being entangled for all bipartitions.
At its core, entanglement transitivity captures the idea that the presence of entanglement in some marginals can be inferred from the entanglement of others---regardless of the specific form that entanglement takes.
Accordingly, the current requirement may be relaxed for a {\em weaker} notion of entanglement transitivity by requiring that the given marginals are not fully separable.
Alternatively, one may strengthen the condition by requiring the marginals---either given or target---to be genuinely multipartite entangled~\cite{Guhne08_EntReview} instead. These are all interesting alternatives that deserve further exploration.

While~\cref{lab: ent_trans_equal_dim} affirms the conjecture proposed in~\cite{Tabia22}, it does not fully account for the findings reported therein.
This is because entangled states do not necessarily violate the positive-partial-transpose (PPT) criterion~\cite{Peres96_PPT, Horodecki96_PPT}, 
and thus~\cref{lab: ent_trans_equal_dim} does not exclude the possibility of finding entanglement transitivity instances with compatible marginals in AC being PPT entangled.
Nonetheless, no such instances have been found from the numerical observations presented in~\cite{Tabia22}, and further numerics presented in this work
(\hyperlink{app:ent-trans-unspecified}{Appendix F},~\cref{tab:ent_trans_not_specified}) again draw the same conclusion.
It therefore remains an open question whether such an example actually exists.
We note, however, in all dimensional combinations listed in~\cref{thm: ent_trans} with $d_A = d_C \geq 3$, 
the requirement $d_B \leq (d-1)^2$ for entanglement transitivity to hold implies $d_B < 4d^2$, which, according to~\cite{Aubrun12_PRA, Aubrun12_PPT}, renders the marginals in AC overwhelmingly likely to be NPPT entangled.
We leave these problems for future investigation.

A broader line of research aims at characterizing the whole from the parts,
addressing questions such as what kind of resource is compatible with the given marginals~\cite{Hsieh24_RMP}.
Examples include separable marginals that are only compatible with entangled global states (see, e.g.,~\cite{TKG+07,TKG+09,Wurflinger12_SepMarg, Navascues21_EMP} and references therein), genuine multipartite entangled global states~\cite{Chen14_SepMarg, Paraschiv18_SepMarginal}, nonlocal global state~\cite{Vertesi14_SepMarg}, nonlocal marginals that are only compatible with other nonlocal marginals~\cite{Chen24_NonlocalityTrans}, etc.
Rather than focusing on special families of quantum states with peculiar marginal resource relations, it is clearly of interest to establish general statements valid for any tripartite system as well.
For tripartite pure states, if one marginal (e.g., $\rho_{AB}$) is PPT entangled, then the other two marginals, $\rho_{AC}$ and $\rho_{BC}$, must violate the reduction criterion and are therefore distillable~\cite{Hayashi11_WeakMonogamy}. 
Furthermore, whether the global state is pure or mixed, at most one bipartite marginal can exhibit a quantum advantage in dense coding~\cite{Prabhu13_DenseCoding}.
Additionally, it would be interesting to study when separable marginals are incompatible with {\em any} separable states, which is an instance of meta-transitivity that has resource-theoretic implications~\cite{Tabia24_EB_metaTrans}.
We note that even two separable states with common overlap can be incompatible sometimes~\cite{Carlen13_incomp_sep}.

\hypertarget{app:entmarg-remark}{%
{\bf\em {Appendix A: Remark on Lemma~\ref{lem: entmarg}---}}%
}Let $\R(\rho)$ be $\rho$'s rank.
Then, it is known that a bipartite state $\rho_{AC}$ is distillable~\cite{Horodecki09_EntReview, Bennett96_ent_distillation}, and hence entangled, if $\R(\rho_{AC}) < \max\{\R(\rho_A), \R(\rho_C) \}$~\citep[Theorem 1]{Horodecki03_rank2_boundstate}.
Now, note that the rank of the marginal $\rho_{AC}$ of a generic pure state $\ket{\psi}_{ABC} \in \C^{d_A} \ten \C^{d_B} \ten \C^{d_C}$ is almost surely given by $\R(\rho_{AC})=\min\{d_B, d_Ad_C\}$ (and similar forms can be obtained for $\rho_{A}$ and $\rho_{C}$).
As a result, we thus conclude that
{\em
$\rho_{AC}$ is almost surely distillable~\footnote{
We note that the distillability of Haar-random pure state marginals has also been discussed in~\cite{Hayden06_GenericEnt,Mori25_no_distillable_marginal}.}
if 
\begin{equation}
\label{eq: distillable ent}
    \begin{aligned}
        \min\{d_B, d_Ad_C\} < \max \big\{ &\min\{d_A, d_Bd_C\},\\
        &\min\{d_C, d_Ad_B\} \big\}.
    \end{aligned}
\end{equation}
}
We now prove that~\cref{lem: entmarg} covers every triple $(d_A,d_B,d_C)$ that fulfills the stronger distillability condition in~\cref{eq: distillable ent}, which can be formally stated as:
\begin{lemma}
    \cref{eq: distillable ent}'s validity implies~\cref{lem: entmarg}.
\end{lemma}

Note that the converse does not hold (e.g., when $d_A=d_B=d_C$).
Thus, in certain regimes of local dimensions, \cref{lem: entmarg} can be more useful in detecting entanglement of generic pure state's marginals.
\begin{proof}
    Suppose~\cref{eq: distillable ent} holds.
    Let $L$ and $R$ be the left- and right-hand sides of~\cref{eq: distillable ent}, so that it reads $L<R$.
    First, observe that $d_B<d_A d_C$ must hold.
    If, instead, $d_B \geq d_A d_C$, then $L=d_A d_C$ while $R \leq \max\{d_A,d_C\}<d_A d_C=L$, contradicting~\cref{eq: distillable ent} (i.e., $L<R$).
    Hence, \cref{eq: distillable ent}'s validity implies $d_B < R$.
    Next, observe that at least one of the inequalities, $d_B < d_A$ or $d_B < d_C$, must hold.
    Assuming the contrary ($d_B \geq d_A$ and $d_B \geq d_C$ both holds) yields $R = \max\{d_A,d_C\}\leq d_B$, violating $d_B<R$.
    Therefore, either $d_B\leq d_A-1$ or $d_B\leq d_C-1$ holds.
    Since $d_A, d_C\geq 2$, it follows that $d_B \leq (d_A-1)(d_C-1)$, which is~\cref{lem: entmarg}.
\end{proof}

\hypertarget{app:multi-marg}{%
{\bf\em {Appendix B: Proof of~\cref{thm: multipartite entmarg}---}}%
}Suppose that we partition the $k$-body marginal, which has $k$ qudits, of a generic $N$-qudit pure state into two parties, one with $m$ qudits and the other with $k-m$ qudits, where $1 \leq m \leq k-1$. 
Based on Lemma~\ref{lem: entmarg}, the $k$-body marginal is entangled with respect to this specific bipartition if 
    $
    {(d^{k-m}-1)(d^m-1) \geq d^{N-k}},
    $
which can be rewritten as
\begin{equation}
\label{eq: KS inequality}
    d^{k-m-\epsilon_1} d^{m-\epsilon_2} \geq d^{N-k},
\end{equation}
where $\epsilon_1\coloneqq y(k-m,d)$ and $\epsilon_2 \coloneqq y(m,d)$,
and the function $y(r,d)$ is defined as
    $
    y(r,d) \coloneqq r - \ln(d^r - 1)/\ln(d),
    $
with $\ln$ denoting the natural logarithm.
Taking the base-$d$ logarithm of both sides of~\cref{eq: KS inequality} yields:
\begin{equation}
\label{eq: rewrite marg ent condition}
     k \ge \dfrac{N+\epsilon_1 + \epsilon_2}{2}.
\end{equation}
Next, we want to show that $y(r,d)$ monotonically decreases in both (integer values of) $r$ and $d$. To this end, it is expedient to extend the domain of $r$ and $d$ to $\mathbb{R}^+$ and note that the partial derivatives of $y(r,d)$ are
\begin{equation}
    \begin{aligned}
        &\dfrac{\partial}{\partial r} 
        y(r,d) = -\dfrac{1}{d^r-1},\\
        &\dfrac{\partial}{\partial d} 
        y(r,d) =\frac{(d^r-1)\ln(d^r-1)-d^r \ln{(d^r)}}{(d^r-1)d\left(\ln{(d)}\right)^2},
    \end{aligned}
\end{equation}
which are smaller than $0$ for $r \geq 1$ and $d \geq 2$. Thus, we conclude that $y(r,d)$ is strictly decreasing when we increase $r$ ($d$) with a fixed value of $d$ ($r$) in this region. 
Since $y(1,2) = 1$, we thus conclude that $\frac{\epsilon_1+\epsilon_2}{2}\le1$ for every $1\le m \le k-1$ and $d\ge2$.
Hence, whenever $k\ge\frac{N}{2}+1$ holds, so does \cref{eq: rewrite marg ent condition}.

Note further that if the {\em stronger} inequality $\frac{\epsilon_1+\epsilon_2}{2}<\frac{1}{2}$ holds, the {\em weaker} condition $k>\frac{N}{2}$ also implies \cref{eq: rewrite marg ent condition}, since both $k$ and $N$ are integers.
For $d=2$, since 
$y(1,2) = 1, y(2,2) \approx 0.4150<\frac{1}{2}$, we have $\frac{\epsilon_1+\epsilon_2}{2}<\frac{1}{2}$ if and only if both $\epsilon_1,\epsilon_2\neq y(1,2)$, i.e., $x\notin\{1,k-1\}$.
For $d \ge 3$, since $y(1,3)\approx 0.3691$, $\frac{\epsilon_1+\epsilon_2}{2}<\frac{1}{2}$ always holds.
The proof is thus completed.
\hfill$\square$

\hypertarget{app:ent-trans}{%
{\bf\em Appendix C: Proof of~\cref{thm: ent_trans}---}%
}Consider generic tripartite pure states $\ket{\psi} \in \hs_A \otimes \hs_B \otimes \hs_C$ uniquely determined by two of its marginals $\rho_{AB},\rho_{BC}$, i.e., $\ket{\psi}_{ABC}$ that satisfy the dimensional constraints given in~\cref{lem: tripartite uniqueness}. 
For the two-body marginals of a uniquely determined $\ket{\psi}_{ABC}$ to exhibit entanglement transitivity, all three two-body marginals of $\ket{\psi}_{ABC}$ should be entangled. 
Next, we prove that this is indeed the case whenever any of the conditions given in \cref{thm: ent_trans} holds.

First, if condition \ref{4-1} in \cref{thm: ent_trans} holds, i.e., $d_A=d_B=d_C=2$, then $\ket{\psi}_{ABC}$ is uniquely determined by $\rho_{AB},\rho_{BC}$~\cite{Chen13_tripartite_uniqueness}. 
In this case, it has also been shown in~\cite{Aubrun13_EntThreshold, Ruskai09_GenericEnt} that the marginals are entangled. 

To show that condition \ref{4-2} in \cref{thm: ent_trans} is also sufficient, consider $d_A=d_C=d\ge3$. 
Then, a sufficient condition for $\rho_{AC}$ to be entangled is $d_B \leq (d-1)^2$ as given in Lemma~\ref{lem: entmarg}. 
Similarly, the inequality $d \leq (d_B-1)(d-1)$ guarantees that both $\rho_{AB},\rho_{BC}$ are entangled. 
The two conditions lead to $\frac{2d-1}{d-1} \leq d_B \leq (d-1)^2$. 
Note that if $d \geq 3$, we have $\frac{2d-1}{d-1} = 2 + \frac{1}{d-1} < 3 < 4\le(d-1)^2$.
This guarantees that the interval $[\frac{2d-1}{d-1}, (d-1)^2]$ contains at least one positive integer, which ensures that the bounds on $d_B$ are non-trivial.
Finally, by \cref{lem: tripartite uniqueness}, $\ket\psi_{ABC}$ is uniquely determined by $\rho_{AB},\rho_{BC}$ for any $d_B\ge2$ since $d_A=d_C=d$~\cite{Huang18_tripartite_uniqueness}.
These thus show that condition \ref{4-2} in \cref{thm: ent_trans} is a non-trivial sufficient condition.

It remains to show that condition \ref{4-3} in \cref{thm: ent_trans} is also sufficient.
If $d_A \neq d_C$, by \cref{lem: tripartite uniqueness}, we may impose $d_B \geq \min\{d_A,d_C\}$ to make $\ket\psi_{ABC}$ uniquely determined by $\rho_{AB},\rho_{BC}$~\cite{Chen13_tripartite_uniqueness}. 
Without loss of generality, let ${\min\{d_A,d_C\} = d_A}$.
For $\rho_{AB}, \rho_{BC}$ to be entangled, we can impose the single condition $d_C\le(d_A-1)(d_B-1)$ from \cref{lem: entmarg}. This is because this single condition implies $d_A \leq d_C\le(d_A-1)(d_B-1)\le(d_C-1)(d_B-1)$, meaning that $\rho_{BC}$ is also almost surely entangled by \cref{lem: entmarg}.
Together with the condition that makes $\rho_{AC}$ entangled from \cref{lem: entmarg}, i.e., $d_B\le(d_A-1)(d_C-1)$, we have
\begin{equation}
\label{eq: ent trans case 3}
    1 + \dfrac{d_C}{d_A-1} \leq d_B \leq (d_A-1)(d_C-1).
\end{equation}
Finally, to ensure \cref{eq: ent trans case 3} is non-trivial, we need to check the interval $[1 + \frac{d_C}{d_A-1}, (d_A-1)(d_C-1)]$ contains at least one positive integer.
For this, we need $(d_A-1)(d_C-1)-\left(1+\frac{d_C}{d_A-1}\right) \geq 1$, which can be verified to hold as long as $d_A\ge3$.
Along with the constraint that ensures the uniqueness from \cref{lem: tripartite uniqueness}, which is $d_B \geq \min\{d_A,d_C\} = d_A$ in this case, we have $d_A \geq 3$ and
\begin{equation}
    \begin{aligned}
        {{\rm max}\left\{d_A, 1 + \dfrac{d_C}{d_A-1}\right\} \leq\ d_B \leq (d_A-1)(d_C-1),}
    \end{aligned}
\end{equation}
which thus completes the proof. \hfill$\square$

\hypertarget{app:multi-ent-trans}{%
{\bf\em Appendix D: Proof of~\cref{thm: multi_ent_trans}---}%
}We focus on the large marginals of generic $N$-qudit pure states that uniquely determine the global $N$-qudit states, pure or mixed.
According to~\cref{thm: multipartite entmarg}, such marginals are also almost surely entangled across all bipartitions. Naturally, the notion of multipartite entanglement transitivity considered in~\cref{thm: multi_ent_trans} concerns marginals having the same kind of entanglement.
We first summarize known sufficient conditions ensuring multipartite uniqueness as follows.
\begin{lemma}[\cite{Huang18_tripartite_uniqueness, Jones05_multipartite_uniqueness, Karuvade18_steadystate}]
\label{lem: multipartite uniqueness}
    For a generic $N$-qudit pure state \(\ket{\psi} \in (\C^{d})^{\otimes N}\), the two sets of marginals each uniquely determine the global state \(\ket{\psi}\):
    \begin{enumerate}
        \item\label{8-1} 
        Two specifically chosen
        $(\lfloor \frac{N}{2} \rfloor + 1)$-qudit marginals.
        \item\label{8-2} 
        $\lfloor \frac{N}{2} \rfloor$ 
        specifically chosen
        $(\lceil \frac{N}{2} \rceil +1)$-qudit marginals.
    \end{enumerate}
\end{lemma}

We now explain how these marginals can be chosen to ensure a unique and compatible global state.
Let us start with condition~\ref{8-1}, which can be obtained from~\cref{lem: tripartite uniqueness}~\cite{Huang18_tripartite_uniqueness} by considering $d_A=d_C=d^{\lceil \frac{N}{2}\rceil-1}$ (see also~\citep[Proposition 5.2]{Karuvade18_steadystate}).
Then, for even $N$'s, one can choose any two distinct $(\frac{N}{2}+1)$-qudit marginals that share exactly two common qudits.
For odd $N$'s, one can choose, instead, any two distinct $(\lfloor \frac{N}{2} \rfloor + 1)$-qudit marginals, which overlap exactly at one qudit.

Next, we elaborate on condition~\ref{8-2} from~\cite{Jones05_multipartite_uniqueness}, which concerns the specification of $\lfloor \frac{N}{2} \rfloor$ marginals having generally a larger overlap than those specified in condition~\ref{8-1}. 
Indeed, we start by choosing {\em any} $\lceil\frac{N}{2}\rceil$ qudits from the $N$ qudits.
Then, the required $(\lceil\frac{N}{2}\rceil +1)$-qudit marginals are obtained by combining these $\lceil\frac{N}{2}\rceil$ chosen qudits, separately, with {\em each} of the remaining $\lfloor\frac{N}{2}\rfloor$ qudits.

It is worth noting that the global state cannot be uniquely determined by specifying only up to $\lfloor \frac{N}{2} \rfloor$-qudit marginals~\cite{Jones05_multipartite_uniqueness}. 
Hence, the size of the marginals listed in~\cref{lem: multipartite uniqueness} cannot be further reduced if we want to exhibit multipartite entanglement transitivity via the uniqueness of a compatible $N$-partite state, pure or mixed.

Combining~\cref{thm: multipartite entmarg} and condition \ref{8-1} of~\cref{lem: multipartite uniqueness} shows that two specifically chosen $(\lfloor \frac{N}{2} \rfloor + 1)$-body marginals coming from generic $N$-qudit pure states exhibit entanglement transitivity whenever $d \geq 3$ or that $N$ is even.
However, when $d=2$ and $N$ is odd, \cref{thm: multipartite entmarg} does not guarantee that these marginals are entangled across all cuts.
Similarly, combining~\cref{thm: multipartite entmarg} and condition \ref{8-2} of~\cref{lem: multipartite uniqueness},
we can also use $\lfloor \frac{N}{2} \rfloor$ specifically chosen $(\lceil \frac{N}{2} \rceil +1)$-body marginals as a set of marginals that can exhibit entanglement transitivity.
This thus gives the following formal version of~\cref{thm: multi_ent_trans}:
{\em
Consider a generic $N$-partite pure state $\ket{\psi} \in (\C^d)^{\otimes N}$.
\begin{itemize}
    \item[(i)] If $d \geq 3$ or that $N$ is even, then the following sets of marginals each almost surely exhibit entanglement transitivity:
    \begin{itemize}
        \item two specifically chosen $(\lfloor \frac{N}{2} \rfloor + 1)$-body marginals;
        \item $\lfloor \frac{N}{2} \rfloor$ specifically chosen $(\lceil \frac{N}{2} \rceil + 1)$-body marginals.
    \end{itemize}
    \item[(ii)] If $d=2$ and that $N$ is odd, then only the second set of marginals almost surely exhibit entanglement transitivity.
\end{itemize}}
This concludes the proof. \hfill$\square$

Let us further remark that for the sake of choosing large enough marginals to ensure entanglement for all cuts and uniqueness of the global state, one can clearly choose marginals even {\em larger} than those specified in~\cref{lem: multipartite uniqueness}. 

\hypertarget{app:marginal-threshold}{%
{\bf\em Appendix E: Proof of marginal threshold in~\cref{tab:ent_all_cuts}---}%
}To prove the marginal size threshold at which a $k$-qudit marginal of a Haar-random pure state becomes entangled across all bipartitions using~\citep[Section 7.2]{Aubrun13_EntThreshold}, we follow the arguments used to prove~\citep[Corollary 2.4]{Aubrun13_EntThreshold}~\footnote{
    However, while their proof also establishes that the threshold marks a sharp transition from marginals being overwhelmingly likely separable to overwhelmingly likely entangled, we do not prove that the bound, ${\frac{N-1}{2}- \mathcal{O}(\frac{1}{\log d})}$, plays the same role. 
    Our result does not imply such a sharp transition.
}.

We consider the same setting as in~\cref{thm: multipartite entmarg}:  
an $N$-qudit Haar-random pure state is divided into three parties, A, B, and C, consisting of $k-m, N-k,$ and $m$ qudits, respectively (see~\cref{fig:multi_ent_marg}).
Explicitely, the local dimensions are given by $d_A = d^{k-m}, d_B=d^{N-k},$ and $d_C = d^m$.
According to~\citep[Section 7.2]{Aubrun13_EntThreshold}, the marginal $\rho_{AC}$ of the Haar-random pure state is overwhelmingly likely to be entangled if $d_B$ is smaller than threshold $s_0(d_A, d_C)$, which depends on $d_A$ and $d_C$, and is estimated to be in between the following range:
$
    c' d_Ad_C \min\{d_A,d_C\} \leq s_0(d_A, d_C) \leq C' d_Ad_C (\ln d_Ad_C)^2 \min\{d_A,d_C\},
$ 
where $c'$ and $C'$ are strictly positive constants independent of the local dimensions, and $\ln$ denotes the natural logarithm.
Without loss of generality, we assume $d_A \geq d_C$ (i.e., $m \leq k/2$).
To ensure that $\rho_{AC}$ is very likely to be entangled, it suffices to require $d_B < c' d_Ad_C^2$ (for large $d$).
Substituting the expressions for the given local dimensions yields:
$
    d^{N-k} < c' d^{k+m}.
$ 
Taking the logarithm base $d$ of both sides returns 
$
    \frac{1}{2}(N - m - \log_d c') < k.
$
To ensure that this holds for all $m \in [1,\frac{k}{2}]$, we must require:
$
    k > \frac{N-1}{2}- \mathcal{O}(\frac{1}{\log d}),
$
as stated in~\cref{tab:ent_all_cuts}.
This concludes the proof. \hfill$\square$

\hypertarget{app:ent-trans-unspecified}{%
{\bf\em Appendix F: Dimensional regimes unspecified by Theorem~\ref{thm: ent_trans}~\cref{tab:ent_all_cuts}---}%
}In~\cref{tab:ent_trans_not_specified}, we explore the behavior of bipartite marginals $\rho_{AB}$ and $\rho_{AC}$ across all local dimension triples $(d_A, d_B, d_C)$, 
where $d_A, d_B, d_C \leq 5$ and, without loss of generality, $d_A \geq d_C$ is assumed, that fall outside the conditions specified in~\cref{thm: ent_trans}.
Specifically, we test whether these marginals
\begin{enumerate}
\item exhibit entanglement transitivity in the AC subsystem;
\item uniquely determine the entire ABC system.
\end{enumerate}
For each dimension combination, we draw $1000$ samples from tripartite Haar-random pure states. 
The PPT criterion~\cite{Peres96_PPT} is used to detect marginal entanglement, while entanglement transitivity is certified using the optimization method adopted from~\cite{Tabia22}.
To assess whether a tripartite Haar-random pure state $\ket\psi_{ABC}$ is uniquely determined by its marginals in AB and BC, we consider the following optimization problem, which can be cast as a semidefinite program (SDP):
\begin{align}
    \underset{\rho_{ABC}}{\text{min}} \quad & \bra{\psi} \rho_{ABC} \ket{\psi}_{ABC} \\
    \text{subject to} \quad 
    & \rho_{AB} = {\operatorname{Tr}_C(\rho_{ABC}) = \operatorname{Tr}_C(\ketbra{\psi}_{ABC}),} \\
    & \rho_{BC} = {\operatorname{Tr}_A(\rho_{ABC}) = \operatorname{Tr}_A(\ketbra{\psi}_{ABC}),} \\
    & \rho_{ABC} \succeq 0,\quad
    \operatorname{Tr}(\rho_{ABC}) = 1.
\end{align}
This yields the minimal possible fidelity ${\mathcal{F} = \bra{\psi} \rho_{ABC} \ket{\psi}_{ABC}}$ between the original state $\ket{\psi}_{ABC}$ and any other tripartite state $\rho_{ABC}$, pure or mixed, that shares the same marginals in AB and AC.
If $1 - \mathcal{F} \leq \epsilon$, for a fixed error threshold $\epsilon$ chosen to be $10^{-6}$ here, we then say that $\ket{\psi}_{ABC}$ is uniquely determined by its marginals in AB and BC up to an error $\epsilon$.

We find that marginals of Haar-random pure states containing at most one qubit typically ($\ge 99\%$) exhibit NPPT (i.e., nonpositive under partial transposition) entanglement transitivity in all considered cases.
In contrast, systems involving more than one qubit may fail to exhibit NPPT transitivity in some instances, suggesting that the dimension constraints of~\cref{thm: ent_trans} may also be necessary in this regime.
There are combinations of dimensions that show generic entangled marginals beyond~\cref{lem: entmarg}, suggesting that the bound may not be tight.
We note that obtaining a tighter bound remains an open problem~\citep[Section 3.D]{Ruskai12}.
Furthermore, in all tested cases, the marginals of Haar-random pure states uniquely determine the global state. 
This includes dimension settings where generic pure-state marginals are not known to ensure uniqueness~\cite{Chen13_tripartite_uniqueness, Huang18_tripartite_uniqueness}, which may be of independent interest.

\begin{table*}[ht]
\centering
\begin{tabular}{
|c
|M{0.5cm}M{0.5cm}M{0.5cm}
|c
|M{1.3cm} M{1.3cm} M{1.3cm} M{1.3cm}
 M{1.3cm} M{1.3cm} M{1.3cm} M{1.3cm}|}
\hline

{\scshape $(d_A,d_B,d_C)$}
 & {\footnotesize\scshape AB}
 & {\footnotesize\scshape BC}
 & {\footnotesize\scshape AC}
 & {\footnotesize\scshape Unique}
 & {\scriptsize\scshape N, N $\to$ N}
 & {\scriptsize\scshape N, P $\to$ N}
 & {\scriptsize\scshape P, N $\to$ N}
 & {\scriptsize\scshape P, P $\to$ N}
 & {\scriptsize\scshape N, N $\to$ P}
 & {\scriptsize\scshape N, P $\to$ P}
 & {\scriptsize\scshape P, N $\to$ P}
 & {\scriptsize\scshape P, P $\to$ P}\\

\hline

$(3,2,2)$ & \cmark & \xmark & \cmark & \cmark 
& 0.931 & 0.069 & 0 & 0 
& 0 & 0 & 0 & 0 \\

\hline

$(3,2,3)$ & \xmark & \xmark & \cmark & \cmark 
& 1 & 0 & 0 & 0 
& 0 & 0 & 0 & 0 \\

\hline

$(4,2,2)$ & \cmark & \xmark & \cmark & \cmark 
& 0.746 & 0.254 & 0 & 0 
& 0 & 0 & 0 & 0 \\

\hline

$(4,2,3)$ & \cmark & \xmark & \cmark & \xmark & 0.999 & 0.001 & 0 & 0 & 0 & 0 & 0 & 0 \\
\hline
$(4,2,4)$ & \xmark & \xmark & \cmark & \cmark & 1 & 0 & 0 & 0 & 0 & 0 & 0 & 0 \\
\hline
$(5,2,2)$ & \cmark & \xmark & \cmark & \cmark & 0.588 & 0.412 & 0 & 0 & 0 & 0 & 0 & 0 \\
\hline
$(5,2,3)$ & \cmark & \xmark & \cmark & \xmark & 0.992 & 0.008 & 0 & 0 & 0 & 0 & 0 & 0 \\
\hline
$(5,2,4)$ & \cmark & \xmark & \cmark & \xmark & 1 & 0 & 0 & 0 & 0 & 0 & 0 & 0 \\
\hline
$(5,2,5)$ & \xmark & \xmark & \cmark & \cmark & 1 & 0 & 0 & 0 & 0 & 0 & 0 & 0 \\
\hline
$(2,3,2)$ & \cmark & \cmark & \xmark & \cmark & 0.910 & 0 & 0 & 0 & 0.090 & 0 & 0 & 0 \\
\hline
$(3,3,2)$ & \cmark & \xmark & \xmark & \cmark & 1 & 0 & 0 & 0 & 0 & 0 & 0 & 0 \\
\hline
$(4,3,2)$ & \cmark & \xmark & \cmark & \cmark & 1 & 0 & 0 & 0 & 0 & 0 & 0 & 0 \\
\hline
$(5,3,2)$ & \cmark & \xmark & \cmark & \cmark & 0.996 & 0.004 & 0 & 0 & 0 & 0 & 0 & 0 \\
\hline
$(5,3,3)$ & \cmark & \xmark & \cmark & \cmark & 1 & 0 & 0 & 0 & 0 & 0 & 0 & 0 \\
\hline
$(5,3,4)$ & \cmark & \cmark & \cmark & \xmark & 1 & 0 & 0 & 0 & 0 & 0 & 0 & 0 \\
\hline
$(2,4,2)$ & \cmark & \cmark & \xmark & \cmark & 0.765 & 0 & 0 & 0 & 0.235 & 0 & 0 & 0 \\
\hline
$(3,4,2)$ & \cmark & \cmark & \xmark & \cmark & 1 & 0 & 0 & 0 & 0 & 0 & 0 & 0 \\
\hline
$(4,4,2)$ & \cmark & \xmark & \xmark & \cmark & 1 & 0 & 0 & 0 & 0 & 0 & 0 & 0 \\
\hline
$(5,4,2)$ & \cmark & \xmark & \cmark & \cmark & 1 & 0 & 0 & 0 & 0 & 0 & 0 & 0 \\
\hline
$(2,5,2)$ & \cmark & \cmark & \xmark & \cmark & 0.589 & 0 & 0 & 0 & 0.411 & 0 & 0 & 0 \\
\hline
$(3,5,2)$ & \cmark & \cmark & \xmark & \cmark & 0.995 & 0 & 0 & 0 & 0.005 & 0 & 0 & 0 \\
\hline
$(3,5,3)$ & \cmark & \cmark & \xmark & \cmark & 1 & 0 & 0 & 0 & 0 & 0 & 0 & 0 \\
\hline
$(4,5,2)$ & \cmark & \cmark & \xmark & \cmark & 1 & 0 & 0 & 0 & 0 & 0 & 0 & 0 \\
\hline
$(5,5,2)$ & \cmark & \xmark & \xmark & \cmark & 1 & 0 & 0 & 0 & 0 & 0 & 0 & 0 \\

\hline

\end{tabular}
\caption{
Entanglement transitivity and uniqueness relative frequencies or dimension triples $(d_A,d_B,d_C)$ not covered by~\cref{thm: ent_trans}.
Each row corresponds to a triple $(d_A,d_B,d_C)$, listed under the assumption that $d_A \geq d_C$.
Columns labeled \textsc{AB}, \textsc{BC}, and \textsc{AC} indicate whether the corresponding bipartite marginals are predicted to be entangled (\cmark) or not (\xmark), according to~\cref{lem: entmarg}. 
The \textsc{Unique} column shows whether the tripartite pure state is predicted to be uniquely determined by its marginals $\rho_{AB}$ and $\rho_{BC}$ (\cmark) or not (\xmark), according to~\cref{lem: tripartite uniqueness}.
The remaining columns categorize the entanglement transitivity behavior.
Each table entry reports the observed relative frequency, estimated from $1000$ samples of tripartite Haar-random pure states per dimension setting.
For example, the left side of the arrow showing \textsc{N, P} refers to the case where $\rho_{AB}$ is NPPT (\textsc{N}) and $\rho_{BC}$ is PPT (\textsc{P}).
The right sides of the arrows indicate whether all compatible marginals in AC are NPPT entangled (\textsc{N}), or at least one is PPT (\textsc{P}).
Among all generated states in ABC, each is tested to be uniquely determined by $\rho_{AB}$ and $\rho_{AC}$, up to an error $\epsilon = 10^{-6}$ (see~\protect\hyperlink{app:ent-trans-unspecified}{Appendix F}).}
\label{tab:ent_trans_not_specified}
\end{table*}

\begin{acknowledgments}
{\bf\em Acknowledgement---}We acknowledge helpful discussions with Antonio Ac\'{\i}n, Guillaume Aubrun, Jianxin Chen, Mengyao Hu, Debbie Leung, Takato Mori, Thomas Theurer, Bei Zeng, and Chao Zhang. 
KSC thanks the hospitality of the Institut N\'eel, where part of this work was completed. 
This work was supported by the Leverhulme Trust Early Career Fellowship (``Quantum complementarity: a novel resource for quantum science and technologies'' with Grant No.~ECF-2024-310), the National Science and Technology Council, Taiwan (Grants No. 112-2628-M-006-007-MY4, 113-2917-I-006-023, 113-2918-I-006-001) and the Foxconn Research Institute, Taipei, Taiwan. 
This research was also supported in part by the Perimeter Institute for Theoretical Physics. 
Research at Perimeter Institute is supported by the Government of Canada through the Department of Innovation, Science, and Economic Development, and by the Province of Ontario through the Ministry of Colleges and Universities.
We acknowledge the use of the magnifier icon from Flaticon.com in our figures.
\end{acknowledgments}

\bibliographystyle{apsrev4-2}
\bibliography{GenericEntMargTrans}

\newpage

\section{Supplementary Material}
\label{Supplementary Material}

\subsection{Detailed proof of Lemma~\ref{lem: entmarg}}

\subsubsection{Haar-random pure states}

The set of pure states in $\C^d$ corresponds to the unit sphere, which we denote by ${\mathcal{S}(\mathbb{C}^d) = \{\ket \psi \in \mathbb{C}^d | \braket{\psi|\psi} = 1\}}$. 
Throughout, by Haar-random pure states $\ket{\psi}$ (or generic pure states), defined in~\citep[Definition 4.1]{Collins15_random_matrix}, we mean quantum states that are uniformly distributed on $\mathcal{S}(\mathbb{C}^d)$.
That is, Haar-random pure states are obtained by uniform sampling from $\mathcal{S}(\mathbb{C}^d)$.
We denote the uniform distribution by $\chi_d$, and use the notation ``$\ket{\psi} \sim \chi_d$'' to indicate that the random pure state $\ket{\psi} \in \mathbb{C}^d$ follows this distribution.

One standard representation of Haar-random pure states is through complex {\em standard Gaussian vectors}, whose formulation is provided in~\citep[Appendix A.2]{Aubrun17_AliceandBob} and~\citep[Chapter 2.2]{Bryc95_GaussianVector}.
Complex standard Gaussian vectors are defined as
\begin{align}\label{eq: Gaussian vectors}
    \ket{G} \coloneqq \sum_{i=1}^d a_i \ket{i},
\end{align}
where the coefficients $\{a_i\}_{i=1}^d \subset \mathbb{C}$ are independent and identically distributed complex random variables with standard Gaussian distribution $N_{\mathbb{C}}(0,1)$, a probability distribution on $\mathbb{C}$ with density $\frac{1}{\pi}e^{-\abs{a_i}^2}da_i$, zero mean, and unit variance. 
We refer to the coefficients $\{a_i\}_{i=1}^d$ as complex {\em standard Gaussian variables}.
The coordinates of a standard Gaussian vector are independent standard complex Gaussians with respect to {\em any} orthonormal basis, so $\{\ket i\}_{i=1}^d$ can be taken to be any orthonormal basis of $\mathbb{C}^d$.

A key fact for our later discussions is that Haar-random pure states can be represented by {\em normalized} complex standard Gaussian vectors, since they are uniformly distributed on $\mathcal{S}(\mathbb{C}^d)$ (see~\citep[Proposition 4.2.3]{Collins15_random_matrix} and~\citep[Theorem 4.1.2]{Bryc95_GaussianVector}).
We denote Haar-random pure states by $\ket{H}$, which is obtained from a complex standard Gaussian vector $\ket{G}$ via normalization:
\begin{equation}
\label{eq: Haar-random pure state}
    \ket{H} \coloneqq \dfrac{\ket{G}}{\sqrt{\braket{G|G}}}.
\end{equation}

\subsubsection{Marginals of Haar-random pure states}
Let us consider the marginals of Haar-random pure states, which we refer to throughout as {\em random marginals}. 
For a review of random marginals, see~\citep[Section 4.2]{Collins15_random_matrix} and~\citep[Section 6.2.3]{Aubrun17_AliceandBob}.
Let $\ket{\psi}_{AB} \in \mathcal{S}\l(\C^{d_A} \ten \C^{d_B}\r)$ be a bipartite Haar-random pure state of the form
\begin{equation}
    \ket{\psi}_{AB} = c \sum _{i=1}^{d_A} \sum_{j=1}^{d_B} a_{ij} \ket{i}_A \ten \ket{j}_B,
\end{equation}
where the coefficients $\{a_{ij}\}_{i,j} \subset \mathbb{C}$ are independent and identically distributed complex standard Gaussian variables,
and $c$ is the normalization constant ensuring $\braket{\psi|\psi}=1$.
Then, the random marginal in system A, denoted by $\rho_A \coloneqq \tr_B(\ketbra{\psi}_{AB})$, takes the explicit form:
\begin{equation}
\label{eq: Wishart matrix}
    \begin{aligned}
        \rho_A
        = \abs{c}^2 \sum_{i,i'=1}^{d_A} \sum_{j=1}^{d_B} a_{ij} a^*_{i'j} \ketbra{i}{i'}_A
        = \abs{c}^2 \sum_{j=1}^{d_B} \ketbra{G_j}_A,
    \end{aligned}
\end{equation}
where each $\ket {G_j} \coloneqq \sum_{i} a_{ij} \ket{i}\in {\mathcal{S}\left(\C^{d_A}\right)}$ is a complex standard Gaussian vector defined in~\cref{eq: Gaussian vectors}, since the coefficients $\{a_{ij}\}$ are independent and identically distributed complex standard Gaussian variables.
Note that $\rho_A$ has unit trace. 
Hence, $\abs{c}^{-2} = \sum_{ij} \abs{a_{ij}}^2$.
The unnormalized matrix $\sum_j \ketbra{G_j}$ appearing in~\cref{eq: Wishart matrix} is known as {\em Wishart matrices}~\cite{Collins15_random_matrix}.
By normalizing each $\ket{G_j}$, we can express $\rho_A$ as a convex combination of Haar-random pure states:
\begin{equation}
\label{eq: random marginal}
    \rho_A = \sum_{j=1}^{d_B} p_j \ketbra{H_j}_A,
\end{equation}
where $\ket{H_j} \coloneqq \frac{\ket{G_j}}{\sqrt{\braket{G_j|G_j}}}$ is a Haar-random pure state since it is a normalized standard Gaussian vector (see also~\cref{eq: Haar-random pure state}), and the weights are given by $p_j = \abs{c}^2 \braket{G_j|G_j}$, satisfying $\sum_j p_j = 1$.
This expression reveals that the range of the random marginal $\rho_A$ is spanned by the set of the collection ${\{\ket{H_j}\}_{j=1}^{d_B}}$ of independently sampled Haar-random pure states from ${\mathcal{S}\left(\C^{d_A}\right)}$.
Since these vectors are almost surely linearly independent (if $d_B \leq d_A$), the range of $\rho_A$ is almost surely a subspace of dimension ${\rm min}\{d_A,d_B\}$.
Note that, as mentioned in the main text, we say that an event happens ``almost surely'' if its complement is of measure-zero with respect to the uniform sampling.
Heuristically, one can view the range of $\rho_A$ as a subspace spanned by a constituent of $d_B$ independently sampled Haar-random pure states from $\mathcal{S}\left(\C^{d_A}\right)$.
Such a subspace is referred to as a {\em random subspace}~\cite{Hayden06_GenericEnt, Walgate08_RandomSubspace, Fujii02_Grassmann}.
The discussion naturally applies to bipartite marginals of tripartite Haar-random pure states, as summarized in the following lemma:
\begin{lemma}[~\cite{Aubrun17_AliceandBob, Collins15_random_matrix}]
\label{lem: random marginal random subspace}
    Consider the bipartite marginal $\rho_{AC}$ of a tripartite Haar-random pure state $\ket{\psi}_{ABC}$.
    A collection of $d_B$ independently sampled bipartite Haar-random pure states from $\mathcal{S}\left(\C^{d_A} \ten \C^{d_C}\right)$ span the range of $\rho_{AC}$.
\end{lemma}

\subsubsection{Proof of Lemma~\ref{lem: entmarg}}

Now, we also recall that a {\em completely entangled subspace} of a bipartite system $\C^{d_A} \otimes \C^{d_B}$ is a subspace that exclusively contains entangled states (see, e.g.,~\cite{Wallach00_CES, Bennett99_UPB, Parthasarathy04_CES, Johnston13_NPTsubspace, Demianowicz24_CESreview}). 
An example of this type is the antisymmetric subspace and the orthogonal complement of the subspace spanned by {unextendible product bases}~\cite{Bennett99_UPB}. 
It has been shown that the maximal dimension of a completely entangled subspace is $(d_A-1)(d_B-1)$~\cite{Wallach00_CES, Parthasarathy04_CES}. 
For a comprehensive review, we refer the readers to~\cite{Demianowicz24_CESreview} and the references therein.

The following result states that subspaces spanned by Haar-random pure states are, almost surely, completely entangled subspaces
as long as their dimensions do not exceed the maximum limit of $(d_A -1)(d_B -1)$. 
This was first established in~\citep[Corollary 3.5]{Walgate08_RandomSubspace} via the Parametrized Sard Theorem. 
An alternative proof using algebraic-geometric methods appears in~\citep[Theorem 8.1]{Aubrun17_AliceandBob}. 
It can also be inferred from properties of the symmetric subspace~\citep[Corollary 10]{Harrow13_SymmetricSubspace}.
\\

\begin{lemma}[\cite{Walgate08_RandomSubspace, Aubrun17_AliceandBob}]
\label{lem: generic CES}
    Let $k\leq (d_A -1)(d_C -1)$.
    Then, the subspace spanned by $k$ Haar-random pure states, independently sampled from $\mathcal{S}\left(\C^{d_A} \ten \C^{d_C}\right)$,
    is almost surely a subspace that contains no product states.
\end{lemma}
Although this lemma is stated with respect to the uniform measure over $\mathcal{S}\left(\C^{d_A} \ten \C^{d_C}\right)$, it remains valid for any probability measure that is absolutely continuous with respect to it.
That is, any measure-zero subset under the uniform distribution will also be of measure zero under such a distribution.

Following the proof in~\citep[Proposition 7.1]{Aubrun13_EntThreshold}, \cref{lem: generic CES} implies that random marginals are almost surely entangled. 
We can now prove Lemma~\ref{lem: entmarg}:
\begin{proof}
[Proof of \cref{lem: entmarg}.]
    Recall that $\ket{\psi}_{ABC}$ is a tripartite Haar-random pure state and its marginal $\rho_{AC} = \tr_B(\ketbra{\psi}_{ABC})$ acting on $\C^{d_A} \otimes \C^{d_C}$ can be written in the form of~\cref{eq: random marginal}:
    \begin{equation}
    \label{eq: bipartite random marginal}
        \rho_{AC} = \sum_{j=1}^{d_B} p_j \ketbra{H_j}_{AC},
    \end{equation}
    where $p_j \geq 0, \forall j$ and $\sum_j p_j = 1$.
    $\ket{H_j} \in \mathcal{S}(\C^{d_A} \otimes \C^{d_C})$ are bipartite Haar-random pure states.
    By~\cref{lem: random marginal random subspace}, the range of $\rho_{AC}$ is spanned by $d_B$ independently sampled Haar-random pure states in $\mathcal{S}\left(\C^{d_A} \ten \C^{d_C}\right)$.
    \cref{lem: generic CES} implies that, with probability one, $\rho_{AC}$'s range contains no product states. 
    Finally, from~\citep[Theorem 2]{Horodecki97_RangeCriterion}, we recall that if a bipartite state is separable, then there exists a set of pure product states that spans its range---a separable state's range {\em must} contain some product states.
    Hence, we conclude that $\rho_{AC}$ is almost surely entangled.
\end{proof}

\end{document}